\documentstyle[prc,aps,floats,psfig,twocolumn]{revtex}
\draft

\begin{document}

\title{
{\rm \vspace*{-0.5cm} \hfill Preprint LBNL-39661 \vspace*{+0.5cm}}\\
Probing the quantum-mechanical equivalent-photon spectrum
for electromagnetic dissociation of relativistic uranium
projectiles
}
\author
{
 Th.~Rubehn \thanks{Electronic address: TRubehn@lbl.gov}
}
\address
{
Nuclear Science Division,
Ernest Orlando Lawrence Berkeley National Laboratory, \\
University of California, 
Berkeley, California 94720
}
\author
{
 W. F. J. M\"uller, W. Trautmann 
}
\address
{
Gesellschaft f\"ur Schwerionenforschung (GSI),
Planckstra{\ss}e 1,
D-64291 Darmstadt, Germany
}
\date{\today}
\maketitle

\begin{abstract}
Electromagnetic fission cross sections for the reactions 
U + (Be, C, Al, Cu, In, Au, U) at $E/A$ = 0.6 and 1.0 GeV
are compared to theoretical calculations using recently
proposed quantum-mechanical equivalent-photon spectra.
In contrast to semi-classical calculations, systematically 
lower cross sections are obtained that cannot reproduce the
experimental results.   
Furthermore, we point out that the study of electromagnetic 
fission cross sections or electromagnetic 1-neutron removal 
cross sections alone cannot provide unambiguous information 
on the excitation of the double giant dipole resonance.
\end{abstract}
\pacs{PACS number(s): 
      25.75.-q,	    
      25.85.Ge,	    
      27.90.+b}	    

\narrowtext

The availability of relativistic heavy-ion beams has enabled 
systematic studies of electromagnetic excitation processes
in nuclei \cite{Eml94}.
Generally, the electromagnetic interaction 
between the projectile and the target nuclei is described 
by the exchange of virtual photons. 
Due to the almost quadratic dependence on the charge of the 
reaction partner and due to the fast time variation
of the Lorentz-contracted electromagnetic field, electromagnetic
cross sections are rather large for 
relativistic heavy-ion collisions;
projectile energies of $\sim$1 GeV/nucleon allow for an 
effective excitation of the giant resonances (10 - 30 MeV). 
When the nucleus is excited above its particle emission threshold,
or, in the case of fission above its fission threshold, it may
then dissociate  according to the appropriate branching ratio. 
Experimentally, measurements have investigated various decay 
branches, like $\gamma$-rays, neutron emission, and fission, 
see, e.g., 
Refs.~\cite{Eml94,Rit93,Sch93,Waj94,Mer88,Hil88,Hil89a,Hil89b,Hil91,Aum93,Aum95a,Aum95b,Jai84,Gre85,Jus94,Pol94,Rub95,Rub96,Arm96,Bor96}.
Almost exclusively, all the experimental data have been compared
with calculations using the semi-classical Weizs\"acker-Williams
method of virtual photons \cite{Fer24,Wei34,Wil34,Ber88}
which has been shown to give an appropriate description
of the process \cite{Ber88}.

Recently, Benesh, Hayes, and Friar have presented new quantum-mechanical
descriptions of the equivalent-photon spectra for electromagnetic 
heavy-ion collisions \cite{Ben96}. 
This recent work has extended previous studies \cite{Ben93,Ben94} 
by examining the sensitivity of nuclear structure inputs. 
Electromagnetic excitation cross sections are calculated 
using the first Born approximation. 
Finally, a model is presented that gives simple quantum-mechanical 
expressions for the E1 and E2  ``equivalent-photon
spectrum'' which can be used with measured photoabsorption cross
sections in exactly the same way as the usual semi-classical
expression. 
Electromagnetic dissociation (EMD) cross sections for a specific
decay channel $\Psi$ can then -- generally -- be expressed by:
\begin{equation}
\sigma_{\rm EMD}^{\Psi} = \int \Big( 
\sigma_{\gamma,\Psi}^{\rm E1} n^{\rm E1}(\omega)
+ \sigma_{\gamma,\Psi}^{\rm E2} n^{\rm E2}(\omega)
\Big) d\omega,
\label{eq1}
\end{equation}
where $\omega$ is the photon energy, $\sigma_{\gamma}$ is the 
photodissociation cross section and $n(\omega)$ is the photon
spectrum generated by the collision partner. The indices E1 and E2
indicate the multipolarities. 
While most semi-classical calculations make use of a cutoff
parameter, $b_{\rm min}$, in coordinate space to account for 
electromagnetic contributions only, the quantum-mechanical
description introduces a cutoff parameter in momentum space
\cite{Ben96}:
\begin{eqnarray}
q_{\rm max} = 1/b_{\rm min} = \hspace*{5cm}\nonumber\\
\Big[1.34 \Big(A_{\rm P}^{1/3} + A_{\rm T}^{1/3}
- 0.75 (A_{\rm P}^{-1/3} + A_{\rm T}^{-1/3})\Big)\Big]^{-1}
\label{BCV}
\end{eqnarray}
where $A_{\rm P}$ and $A_{\rm T}$ are the mass numbers of the
projectile and target, respectively \cite{Ben89}. 
It has been shown that the used parameterization of $b_{\rm min}$ 
allows for a good description of the total nuclear reaction cross 
section \cite{Hes96}.

In this paper, we will not discuss the quantum-mechanical ansatz 
{\it per se} which has been presented in Ref.~\cite{Ben96};
this issue will be addressed elsewhere \cite{Ber96}.
However, we shall apply the given quantum-mechanical virtual photon
spectra to calculate electromagnetic dissociation cross sections 
of relativistic heavy-ion collisions. Comparisons between 
these calculations and semi-classical calculations on one 
hand and experimental results on the other hand will be 
discussed. In particular, it will be shown that due to
uncertainties in various input parameters, {\it no} 
unambiguous conclusions on the excitation of the double giant dipole 
resonance (DGDR) can be drawn by {\it only} investigating the 
electromagnetic 1n removal cross sections 
as it has been done in Ref.~\cite{Ben96}. 

\begin{figure}[htb]
 \centerline{\psfig{file=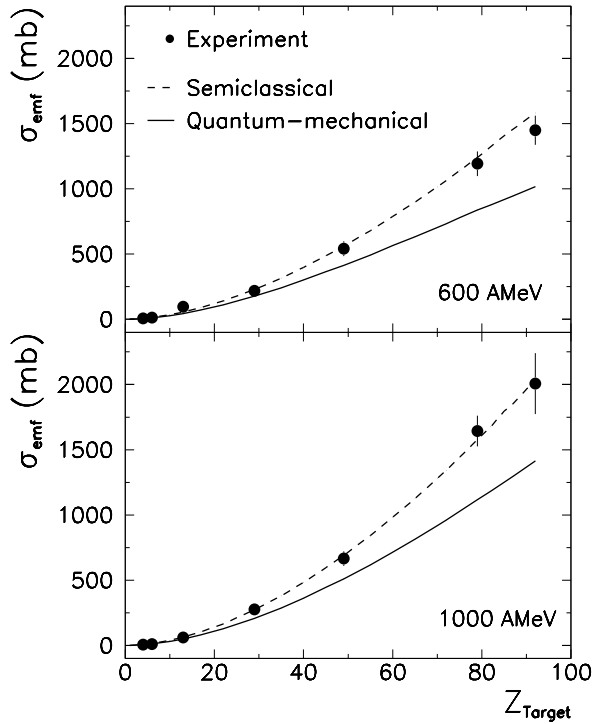,height=10.0cm}}
 \caption[]{
 Electromagnetic fission cross sections for the reactions
 $^{238}$U + (Be, C, Al, Cu, In, Au, U) at E/A = 600 and 1000 MeV
 (from Ref.~\protect\cite{Rub95}).
 For comparison, theoretical results are shown using both the 
 semiclassical (dotted line) \protect\cite{Rub95} and the quantum 
 mechanical descriptions (full line) \protect\cite{Ben96} 
 of the photon spectra.
 }
 \label{xs}
\end{figure}

We will first focus on the comparison of EMD cross sections. 
In a previous work, electromagnetic fission (EMF) of $^{238}$U projectiles 
has been experimentally studied \cite{Rub95} using the ALADIN spectrometer 
at the heavy-ion synchrotron at GSI, Darmstadt. Seven different targets
(Be, C, Al, Cu, In, Au, and U) have been bombarded at  0.6 and 1.0 
GeV/nucleon. Experimental details on the 
measurements and the setup can be found in Refs.~\cite{Rub95,Rub96b,Sch96}.
In Fig.~\ref{xs}, the deduced electromagnetic fission cross 
sections taken from Ref.~\cite{Rub95} are shown as a function of the 
charge number of the target for both bombarding energies. 

For comparison, both the semi-classical and the quantum-mechanical 
calculations have been performed in the same fashion using Eq.~\ref{eq1}.
The use of parameterizations of the photodissociation cross
sections $\sigma_{\gamma, \rm total}$ \cite{Vey73}, the fission 
probability and the cutoff parameter where nuclear interactions 
become dominant \cite{Ben89} has already been discussed previously 
\cite{Rub95}.

While the semi-classical calculations can well reproduce the 
experimental data, the quantum-mechanical calculations give 
significantly lower cross sections. 
This is due to a reduced photon flux in the quantum-mechanical
description. 
Note that this discrepancy is larger than the uncertainties 
connected with the choice of the cutoff parameter or the 
resonance parameters (see, e.g., Refs.~\cite{Rub95,Rub96}):
the Kox parametrization \cite{Kox} of the cutoff parameter results  
in $\sim$15\% lower cross sections, even the use of 
photodissociation cross sections \cite{Cal80} which are known 
to be systematically high compared to other measurements 
\cite{Wil90,Rie90,Web90,Gur76}, results in only 15\% higher 
cross sections.
The experimental data can only be reproduced by the 
quantum-mechanical calculations when the cutoff parameter
in momentum space $q_{\rm max}$ is artificially increased
by 30\%. For the reaction $^{238}$U + $^{238}$U, this
results in a value of $b_{\rm min}$ = 12.5 fm compared
to $b_{\rm min}$ = 16.3 fm from Eq.~\ref{BCV}. 
This significantly lower value seems inappropriate in
terms of the total absorption radius and is not in
agreement with previous works investigating total
reaction cross sections (see, e.g., Ref.~\cite{Hes96}).
While the equivalence between the sharp and smooth cutoff
$b_{\rm min}$ has been shown to be valid in coordinate space
\cite{Aum95a}, it seems to be invalid for the cutoff 
$q_{\rm max}$ in momentum space \cite{Ber96}.

Furthermore, calculations predict nearly the same cross sections for 
electromagnetic fission, whether or not the possibility
of two phonon excitation is included.
This is due to the fact that the higher fission
probability in the energy regime of the DGDR is to a large
extent compensated by the redistribution of cross section
from 1-phonon to that of 2-phonon excitation \cite{Rub95}. 
Therefore, the results of the calculations are almost 
independent of the strength of the DGDR as long as 
EMF cross sections are discussed. 
The comparison of calculated EMF cross sections with experimental
data is thus {\it virtually independent} of the excitation of the DGDR.
We conclude that the quantum-mechanical photon spectra
presented in Ref.~\cite{Ben96} are -- in contrast to semi-classical
photon spectra -- not able to describe the 
experimental results.

Electromagnetic fission cross sections alone do
not allow one to draw conclusions on the excitation of the 
DGDR. This is also true for the 1-neutron removal cross
sections measured in 
Refs.~\cite{Mer88,Hil88,Hil89a,Hil89b,Aum93,Aum95a,Aum95b}.
It has been shown previously that the 1n cross sections 
calculated using multiphonon excitations of the GDR
differ only insignificantly from simple calculations 
based upon the excitation of the 1-phonon state only \cite{Aum93}.

Many efforts have been made to measure quantities which 
permit a more direct test of the excitation of the 
double GDR \cite{Rit93,Sch93,Waj94,Aum93,Aum95b,Rub95,Bor96}.
In the following, we briefly review some of these observables:

1) The asymmetry of the fission fragment charge distribution
is known to be very sensitive to the excitation energy distribution.
The asymmetry is usually expressed by the 
peak-to-valley ratio of the double humped charge
distribution. 
In two independent experiments, a peak-to-valley ratio of 
7.6$\pm$2.6 and 7.1$\pm$1.0, respectively, has 
been found \cite{Rub95,Arm96}. Calculations show that
the excitation of the single phonon states alone would
result in a significantly higher peak-to-valley ratio of 
16$\pm$3, while calculations which account for the
excitation of the DGDR can reproduce the experimental
findings.
Therefore, the low peak-to-valley ratio has been 
interpreted as a clear evidence of the DGDR excitation. 
This conclusion is also supported by the measurement of 
the proton odd-even effect of the fission fragment distribution
\cite{Rub95}.
We note that these quantities are completely independent
of the integrated EMF cross sections.

2) Aumann {\it et al.} have studied electromagnetic
dissociation by measuring 1n-5n neutron removal cross
sections for various reactions \cite{Aum93,Aum95b}. 
Studying the 2n, 3n, 4n, and 5n removal cross sections allows 
for a sensitive test of the contributions from higher 
excitation energies.
It has been shown that these data cannot be understood 
without the excitation of the DGDR which accounts for the
largest fraction of the cross section in the competing 
channels \cite{Aum93,Aum95b}.

3) Recently, Boretzky {\it et al.} have reported conclusive 
evidence for the excitation of the DGDR in $^{208}$Pb 
\cite{Bor96}. By measuring both $\gamma$-rays and neutrons,
the excitation energy was reconstructed and differential
cross sections d$\sigma$/d$E^*$ were deduced for several
targets. These results allow for the comparison 
of the differential cross sections as a function of excitation
energy with calculations and provide a direct test of the 
strength of the DGDR. 

The experimental evidence listed above contradicts the 
theoretically deduced statement by Benesh, Hayes, and 
Friar \cite{Ben96} saying that the quantum-mechanical 
photon spectra leave little room for multiphonon mechanisms.


In conclusion, we have applied the quantum-mechanical
equivalent photon spectra presented in Ref.~\cite{Ben96}
in order to make comparisons with experimental results.
It turns out that electromagnetic fission cross sections 
of uranium in the energy regime of $\sim$1 GeV/nucleon
are independent of the strength of the DGDR and thus 
permit tests of theoretical calculations without 
complications that may result from multiphonon excitations.
Comparisons with experimental electromagnetic fission data
show that, due to the significantly lower photon flux
of the quantum-mechanical description, the 
experimental EMF cross sections cannot be described by
the calculations. Since uncertainties in the parametrizations
are significantly smaller, we conclude that the quantum-mechanical 
photon spectrum presented in Ref.~\cite{Ben96} is not able to 
describe the experimental results.
Furthermore, no conclusive information on the excitation of
multiphonon mechanisms can be drawn from the study of 
cross sections of electromagnetic fission or 1n removal only.

\bigskip    
This work was supported by the Director, Office of Energy Research,
Office of High Energy and Nuclear Physics, Nuclear Physics Division
of the US Department of Energy, under contract DE-AC03-76SF00098.


\end{document}